# Cryptocurrency Smart Contracts for Distributed Consensus of Public Randomness

Peter Mell[1], John Kelsey[1,2], and James Shook[1]

[1] National Institute of Standards and Technology, Gaithersburg MD, USA

[2] Department of Electrical Engineering, ESAT/COSIC, KU Leuven, Belgium

**Abstract.** Most modern electronic devices can produce a random number. However, it is difficult to see how a group of mutually distrusting entities can have confidence in any such hardware-produced stream of random numbers, since the producer could control the output to their gain. In this work, we use public and immutable cryptocurrency smart contracts, along with a set of potentially malicious randomness providers, to produce a trustworthy stream of timestamped public random numbers. Our contract eliminates the ability of a producer to predict or control the generated random numbers, including the stored history of random numbers. We consider and mitigate the threat of collusion between the randomness providers and miners in a second, more complex contract.

## 1 Introduction

Most modern computing devices can produce secure random numbers. However, there are applications which require that many parties share and trust some source of random numbers. For example, running a lottery requires some trustworthy source of *public random numbers*. In the rest of the paper, we define a lottery abstractly as any mechanism that randomly picks a proper subset of elements from some larger set. It is necessary to ensure that the chosen subset cannot be predicted (before some published time), controlled (deliberately set), or influenced (biased toward values that are more desirable for some party). The interesting research question is: *how can we get trustworthy public random numbers sampled from a uniform distribution, especially when the producer of random numbers has a financial incentive to cheat?*

Currently an individual 'beacon' service, a public producer of randomness, may use specialized hardware setups and cryptography to reduce the possibility of the numbers to be compromised [3]. However, the ability to control the numbers (by the beacon owner or some attacker that has compromised the beacon) may remain. What is needed is a consensus protocol for a set of mutually distrusting entities to collaborate to produce a trustworthy stream of publicly available random numbers.

Our solution is to create an Ethereum[1] [22] smart contract, called a *lighthouse*, which implements a beacon service while taking as input random numbers from one or more external and potentially malicious randomness producers. To produce the lighthouse output, we combine producer input with blockchain hashes while forcing producers to commit to future values. In creating the distributed consensus protocol, we leverage the security capabilities associated with smart contracts and blockchains along with a novel commitment system we call Merlin chains (which mitigates a vulnerability common in other systems). Our lighthouse service's timestamped random outputs are published on the Ethereum blockchain, which ensures their immutability and their public visibility. This merging of beacons, smart contracts, and blockchains enables the production of public random numbers at an extremely high level of security, even when assuming the presence of powerful malicious actors in the system (as long as all participating actors aren't malicious).

We provide two main proposed designs:

1. A **single-producer contract** which provides security against control or influence from the randomness producer *or* a large coalition of miners competing in the digital currency system, but not against both.
2. A **multiple-producer contract** which provides security against control or influence from all $k$ of the randomness providers colluding, or a large coalition of miners conspiring with $k-1$ of the randomness providers.

Both designs publish random numbers along with a time before which the random number could not have been predicted by any entity, thus eliminating prediction attacks. With these designs, we have provided a solution for the trustworthy public production of streams of immutable public random numbers. Finally, we create such a contract and empirically test it on the Ethereum test network using both the single and multiple producer models.

Usage of lighthouse services can greatly benefit any public lottery so that selection of random numbers is no longer done behind closed doors, where the public has to trust that no cheating is taking place. Lotteries enable a limited set of resources to be fairly chosen for, or distributed to, a set of customers. Among many other areas, their uses include school placements, dorm rooms allocations, gambling, military drafts, jury duty, immigration applications, election site auditing, and large public financial games run by governments. The utility of a beacon extends far beyond lotteries, but a complete discussion of those applications is outside the scope of this paper.

Different types of public lotteries are more or less sensitive to the three attack types mentioned previously: prediction, control, and influence. For example, with election site auditing an attacker primarily wants to ensure that the election sites chosen for auditing do not correspond to the compromised sites. The attacker then primarily wants *influence* to change the sites chosen for audit if the unmodified result is going to include a compromised site. However, in a gambling scenario, the attacker probably wants to *predict* the winning number or, even better, *control* the result. Our approach must mitigate all three types of attack.

---

[1] Any mention of commercial products is for information only; it does not imply recommendation or endorsement by NIST.

The rest of this paper is organized as follows. Section 2 discusses previous and related work. Section 3 discusses background information. Section 4 provides partial solutions that build towards our final solution. Section 5 describes our design for a single producer contract. Section 6 describes our multiple producer contract; Sect. 7 discusses our empirical work; and Sect. 8 concludes.

## 2   Previous and Related Work

The original idea of a beacon (a public service that publishes signed, timestamped random numbers) comes from Rabin [16]. More recently, in [11], Fischer et. al. propose the usefulness of a beacon service, and describe the NIST beacon. They also propose a general protocol to allow many beacons to be used together to decrease required trust in a single TTP/point of failure, and describe some practical applications for a beacon service. There have also been many attempts to find verifiable public random numbers for use in other applications, such as election auditing [10] and the choice of parameters in cryptographic standards [7].

The simplest way to build a beacon is to simply set up a trusted machine, which generates and signs timestamped random numbers. Existing services such as the NIST Beacon [3] and the beacon-like random.org [6] follow this approach. For many applications of a beacon, this provides sufficient practical security. However, it has a single point of failure – the owner of the beacon (or anyone who compromises the trusted machine on which the beacon is running) can influence or predict future random numbers[2].

### 2.1   Entropy from the Environment

In order to avoid a single point of failure or trust, many people have tried to use unpredictable data from the world to generate public random numbers. In order to be useful, these numbers need to be public, widely-attested, and not under anyone's control.

In [10], the authors consider using financial data as a source of randomness, particularly for election auditing, and use existing tools from finance to estimate the entropy and difficulty of influencing these numbers. [7] considers the use of public financial lotteries to generate random numbers (intended for use in defining cryptographic standards). [8] uses the hash of a block from the Bitcoin blockchain and analyzes the cost of exerting influence on these random numbers by bribing miners to discard inconvenient mined blocks. Our approach uses block hashes in a related way and we have to consider similar attacks.

---

[2] The NIST Beacon's published format includes features to mitigate some attacks–for example, the beacon operator cannot directly control the beacon outputs, as they're the result of a SHA512 hash. However, he *can* predict and influence future random numbers.

## 2.2 Combining Randomness from Multiple Parties

Still another approach is to combine random values from multiple sources, with the goal of getting a trustworthy public random number if enough of the contributors are honest. This may be done by first collecting *commitments* from participants[3], and then asking each participant to *reveal* their commitments.

For example, if Alice and Bob want to each furnish a part of a shared random number, Alice generates random number $R_A$ and publishes hash($R_A$), while Bob generates $R_B$ and publishes hash($R_B$). After both commitments are published, Alice and Bob reveal their random numbers, and agree to use $R_A \oplus R_B$ as their shared random number. (This is referred to as a commit-then-reveal protocol.) The generic attack against this kind of scheme is for Alice to wait until Bob has published $R_B$, and then decide whether she likes the resulting random number or not. If not, she can "hit the reset button," claiming to have suffered a system failure that caused her to lose $R_A$. If this leads to the shared random number being generated again in an actual random way (even in a way that excludes Alice), she has now exerted some influence on the shared random number.

*Commit-Then-Reveal Approaches* The new NIST Beacon format [12] has a precommitment field intended to allow for combining of beacons using a committhen-reveal protocol. However, preventing the 'hit the reset button' attack is left to be handled by reputation–a beacon that skips providing an output often will get a reputation for unreliability. The Randao [4] is an Ethereum service that tries to solve this problem by requiring each party that contributes a commitment to also post a performance bond. Anyone who refuses to reveal their random number forfeits the bond. [19] describes an elaborate set of protocols to use verifiable secret sharing and Byzantine agreement to generate public random numbers from $3k$ independent participants, so that the shared random numbers will be trustworthy (and impossible to prevent from being published) so long as at least $k + 1$ participants are trustworthy.

*Variants Using Slow Computations* [13] takes a different approach to combining contributions from multiple parties. Contributions from the public as well as environmental inputs from a public video camera are hashed together and the hash is published. The inputs are fed into an inherently sequential computationally slow hash function, and much later after the hash is computed the result is published. Since nobody could have known the result of the slow hash function when the inputs were hashed and published, nobody could have influenced the output by deciding what or whether to send an input in. A related approach is considered in [9], in which a computationally slow function is used to produce shared random numbers from Bitcoin or Ethereum block hashes while preventing miners from influencing the resulting random numbers. The same paper describes a set of protocols for ensuring that the computationally slow function is correctly computed, and considers the necessary financial rewards for incentivizing participants to keep verifying the correctness of the computation. Another related possibility to prevent an attacker "hitting the reset button" is to use time-lock puzzles, as described in [17]. If Alice publishes $TL(R_A)$, where $TL()$ is a time-lock scheme with a minimum time to unlock of one hour, and then five minutes later all parties reveal their random numbers, the attack is prevented. Even if Alice wants to hit the reset button (refuse to publish her number to stop the beacon from publishing), she can only delay knowledge of the shared random number for one hour.

---

[3] Without these commitments, Alice can always wait for Bob to publish a random number, and then choose hers to control the resulting shared value.

*Merlin Chains* In this paper, we describe still another approach, called a Merlin chain, to address this problem by giving participants a way to credibly commit to being able to recover their 'lost' random numbers after hitting the reset button. This is an example of a common situation, in which a party in a protocol becomes more capable by restricting its future freedom of action[4].

## 3  Background

Beacons are entities that produce a stream of random numbers [16] (see [3] for a currently-operating example). Each time a beacon releases a random number, it is called a 'pulse'. Beacons have three properties:

1. A beacon will put a random number $R$, unpredictable to anyone outside the beacon itself, in each message.
2. A beacon will never release a signed random number with a timestamp $T$ before time $T$ (so nobody outside the beacon could have known the random number earlier than that time).
3. A beacon will emit only one random number for each timestamp $T$.

In order to be useful, the outputs from a beacon must be publicly available and must be immutable. A beacon pulse may have many fields, but only two are really essential: the random number, $R$, and the timestamp, $T$.

---

[4] A more general version of this idea appears in [18], applied to many real-world situations that can be modeled by game theory.

Blockchains are immutable digital ledger systems and were first used for digital cash with Bitcoin [15]. Each 'block' contains a set of transactions as well as the hash of the previous block (thus forming the 'chain'). They can be implemented in a distributed fashion (without any central authority) and enable a community of users to record transactions in an immutable public ledger. This technology has undergirded the emergence of cryptocurrencies where digital transfers of money take place in distributed systems; it has enabled the success of currencies such as Bitcoin [15] and Litecoin [2]. In such systems, a community of 'miners' maintain a blockchain by competing to solve a mathematical puzzle. The solution is evidence that the miner is performing computation, and for this reason such system are called 'proof-of-work' systems. The 'miner' that solves the current puzzle can then publish the next 'block' which contains recent digital cash transactions. The winning miner receives a block award and may receive fees from included transactions, both in terms of the applicable electronic currency. Some blockchains use other techniques, such as consensus among trusted nodes, proof-of-stake, or proof-of-storage. Without modification, our protocol will work only with 'proof-of-work' systems.

Ethereum [22] is a blockchain-based cryptocurrency that supports 'smart contracts'. Contracts are programs whose code and state exist on the public blockchain and they can both send and receive funds while performing arbitrary computations. They can act as a trusted third party in financial transactions, since the code is public but immutable. The programming language used for contract transactions, Solidity [5], is limited in functionality but is Turing Complete [20]. Ethereum charges a fee for contract execution, called 'gas'. The originator of any transaction must pay this fee or the transaction aborts. There is a maximum gas limit, currently 3000000, to prevent computationally expensive programs from being submitted to the Ethereum miners (since each miner will execute each transaction in parallel).

### 3.1 Merlin Chains

In the rest of this paper, we use a sequence of unpredictable numbers we call a *Merlin chain*[5]. This is a (usually long) sequence of values where every value $V_x$ is the hash of the value with the next higher index $V_{x+1}$ (i.e., $V_x$ = SHA3($V_{x+1}$)). This use of a hash function then provides a series of random values taken from a uniform distribution but where each value is related to the previous value (because the current value is created by hashing the previous value).

A Merlin chain has three important properties:

---

[5] The Merlin Chain is named after the character of Merlin in T.H. White's *The Once and Future King*[21], who lives his life backwards in time.

1. An attacker who has seen all previous entries ($V_{0,1,2,\ldots,j-1}$) in the Merlin chain cannot predict anything about the next entry ($V_j$).
2. Each entry in the chain works as a *commitment* to the next entry in the chain. Once an entity has revealed $V_0$, it has no valid choice except to follow this with $V_1$, then $V_2$, and so on.
3. By storing $V_n$ offsite, the entity revealing the chain entries can guarantee that even a catastrophic hardware or software failure will not prohibit the production of chain values (as would happen were the chain data lost).

The most important feature of the Merlin chain is that it takes away the choices of the entity using it, while still allowing that entity to produce numbers (unpredictable to everyone else). For the user of the Merlin chain, "Everything not forbidden is compulsory"[21].

## 4 Preliminary Approaches

In this section, we describe some plausible-sounding strategies to make a beacon. These approaches don't work but will build towards our proposed solution, thus motivating our design choices in the rest of the paper.

### 4.1 Block hashes

Each block in the Ethereum blockchain is hashed using 256-bit SHA3 and this result is published on the blockchain along with a timestamp. This meets our definition of a beacon in Sect. 3 and one might consider using these hashes as a source of public randomness. However, in this case it turns out that it is possible for the Ethereum miners to influence the beacon results. Consider the situation where a coalition controlling a fraction $F$ of all the processing power of the Ethereum miners is working to predict, control, or influence a block hash. Predicting the block hash would require knowing all transactions to be included in the blockchain up to and including the block whose hash will be used for a random number. Thus, prediction a very short time in advance is sometimes possible for a coalition of miners but prediction far in advance would require control of the whole mining pool and a very visible-to-the-world denial of service attack on the transactions submitted to Ethereum. With respect to control, it's clear that even when $F$ = 100%, there is no way for the coalition to control the value of the block hash, since it's the output of a hash function.

However, influencing the block hash is quite feasible. Consider a coalition controlling $F$ % of the total mining power, which wants to force a single bit of the block hash to be a one. The coalition members attempt to mine the next block, but when they reach a valid proof of work (so that they've successfully mined a block) they check to see whether the resulting block hash has the desired bit set. If not, they simply throw the block hash away and keep trying to mine the next block.

Table 4.1 shows the result of simulating this attack, for various fractions of mining power controlled by the coalition.

**Table 1.** Extent to which a coalition of miners can influence one bit of the block hash

| Fraction of processing power in coalition | Bias in targeted bit |
|---|---|
| 5% | 0.01 |
| 10% | 0.03 |
| 20% | 0.06 |
| 30% | 0.09 |
| 40% | 0.13 |
| 50% | 0.17 |

As the table shows, even a coalition with only 10% of the miners' processing power can impose a potentially significant amount of bias on a selected bit of the block hash, causing the selected bit to have probability 0.53 of being a one.

**4.2 Adding a Producer of Randomness**

The above analysis demonstrates why the block hash alone cannot be used as a public source of randomness. We now consider adding an external producer of randomness, moving us closer to a useful solution. The producer sends a random number $V$, and then the contract produces an output $R = SHA3(H \Vert V)$, where $H$ is the block hash of the previous block. If the producer does not reveal $V$ until the block hash is calculated, the miners no longer can exert any influence over $R$. However, in this scenario the producer can choose $V$ after $H$ is generated and thus influence $R$. In addition, this influence is greater since it is very easy for the producer to compute many $R$ values by simply changing the $V$ input (it is much harder for the miners because to compute a new candidate $R$ value they must create a blockchain block that wins the current block competition).

Our solution to these residual security issues is for the contract to require the producer to generate $V$ prior to $H$ being computed. It does this by requiring that the producer submit the hash of $V$ before it records the value of $H$ to be used. Then only after $H$ is computed by the miners, the producer submits $V$ to the contract. The contract can check that this is the value the producer committed to upfront by simply hashing $V$. The miners can't influence $R$ because they don't know $V$ when computing the block hash. The producer can't influence $R$ because it can't know the block hash when initially committing to a $V$ value (when it sends the hash of $V$ to the contract). The next sections more formally present this approach and handle a variety of security issues that arise (including the possibility that the producer and miners might collaborate to circumvent the security architecture).

## 5 Single Producer Contract

In this section we present a contract whose input comes from a single producer and whose output is a beacon. It is designed to produce a 32-byte random number on the blockchain with a maximum frequency of about once every 30 seconds (more precisely once every other Ethereum block). To maximize the usability of the provided beacon service, we recommend that the producer provide input to the contract at some fixed interval greater than 30 seconds.

The producer will provide unpredictable values from a Merlin chain, and so must pre-compute all inputs that will be provided to the contract for its lifetime. Let $n$ represent the chosen number of input values. The value $V_n$ is chosen randomly, $V_{n-1}$ = SHA3($V_n$), $V_{n-2}$ = SHA3($V_{n-1}$) and so on until the computation of $V_1$. The Merlin values are released to the contract starting with $V_1$ (the reverse of the order in which they were generated).

The function `B()` will provide the block number in which some input or output is processed by the contract. The function `BH()` provides the block hash of some block number. Lastly, the function `timestamp()` provides the Ethereum timestamp for some block.

The producer will periodically provide the contract some message containing a $V_x$ value along with a timestamp $U_x$. The contract in response may produce a random value $R_x$ and a timestamp $T_x$ (note that in certain circumstances the contract may not publish an $R_x$ value). $T_x$ will be the time before which no entity could have predicted $R_x$, including the producer (usually this will be about 30 seconds prior to $R_x$ being publicly released).

The core idea is that for each message (containing some $V_x$) received from the producer, the contract will attempt to generate $R_x$ using as input both an Ethereum block hash and $V_x$. The block hash used will be one that was generated after $V_{x-1}$ was submitted to the contract but before $V_x$ was submitted. This way the miners can't know $V_x$ when the relevant block hash is created and they can't then influence $R_x$ (assuming that the producer and a group of miners are not colluding). Likewise, the producer can't influence $R_x$ because $V_x$ was predetermined by the submission of $V_{x-1}$ and this was done before the relevant block hash was generated. $T_x$ is then generated by taking the minimum of $U_{x-1}$ and the Ethereum timestamp for the block in which $V_{x-1}$ was submitted (taking the minimum eliminates malicious producers from being able claim a Merlin value was revealed later than it was revealed). The actual protocol is slightly more complicated (to account for unexpected input, messages submitted too early, and Ethereum implementation issues). It is outlined below.

## 5.1 Single Producer Protocol

For each message, with associated $V_x$ and $U_x$ values, the contract checks the following prior to accepting the input:

1. The message must come from the Ethereum address registered in the contract as the one pertaining to the producer.
2. $V_x$ must be the next value on the producer's Merlin chain (i.e., $V_{x-1}$ = SHA3($V_x$)). This ensures that the producer can't influence $R_x$.

However, $V_x$ is not considered 'valid' for producing a random number, $R_x$, and a timestamp, $U_x$, unless the following hold (assume that $R_y$ is the last produced $R$ value, usually $R_{x-1}$):

1. The block number in which $V_x$ is processed by the contract must be at least 2 more than the block number where the last valid $V$ value was processed by the contract[6] (i.e., B($V_x$) ≥ B($R_y$)+2). This ensures that the miners can't use the block hash to influence $R_x$ (since miners can discard a block after computing the block hash).
2. The contract must have access to BH(B($R_y$)+1). The contract will retrieve this given any activity (either from the producer or any customer retrieving random numbers) but Ethereum only provides access to the blockhashes for the last 256 blocks. If this is not available[7], the contract will output a public error log message and reset the block hash used to be the one from the next Ethereum block (i.e., BH(B($V_x$)+1).)

If these conditions are satisfied, $R_x$ and $T_x$ are generated according to the following formulas:

$$R_x = \text{SHA3}(V_x \| \text{BH}(B(R_y) + 1)) \tag{1}$$

$$T_x = \min(\text{timestamp}(B(R_y)), U_x) \tag{2}$$

Figure 5.1 provides an example of two valid messages arriving to the contract and shows how the contract uses them to generate $R$ and $T$ values. In the figure, we use $b_x$ to represent the block number at which some $V_x$ arrived to the contract.

---

[6] The producer can ensure this is always true by verifying that it doesn't send the next ($V_x, U_x$) message until it has seen at least one block go past on the blockchain since the last random output.

[7] This availability could be ensured by setting up another provider which does nothing except send a message to the lighthouse contract once every 256 blocks (since blockhashes produced more than 256 blocks in the past are irretrievable in the Ethereum system).

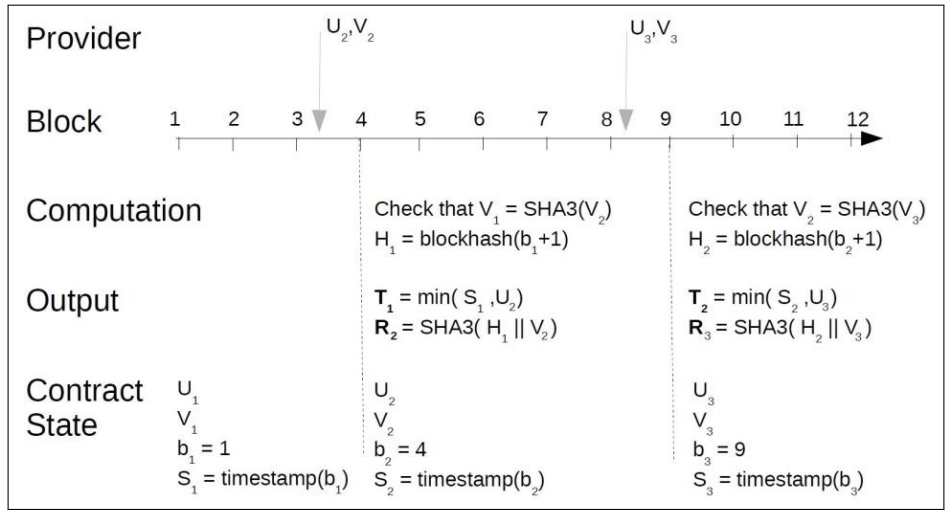

**Fig. 1.** The Single Producer Protocol

### 5.2 Mitigated Security Flaws

We now analyze different attack scenarios and discuss how they are mitigated:

1. The producer might try to use $V_x$ to influence $R_x$. However, this won't work because $V_x$ is fixed based on $V_{x-1}$ and the block hash used was generated after $V_{x-1}$ was revealed.
2. The producer might try to delay sending $V_x$ to influence $R_x$. This was possible in earlier designs where the block hash used for $R_x$ was the one prior to $V_x$. In this case, the producer could watch the block hashes being produced and then quickly issue a pulse after a desirable block hash was published on the blockchain. We mitigated this by fixing the block hash to be used to be $BH(B(R_y)+1)$.
3. A producer could purposefully submit a message too early. However, the message is rejected as invalid and this simply updates the Merlin value $V$ stored on the contract (which is fine since the relevant block hash has not yet been generated).
4. Because of a design limitation in the Ethereum Solidity language, the contract is only able to retrieve up to the last 256 block hashes (about 68 minutes of blockchain operation). The threat is that prior to revealing $V_x$, a producer might calculate $R_x$ and find it undesirable. The producer may then wait 256 blocks prior to releasing $V_x$ so that the correct blockhash can't be retrieved. This effectively changes the result since the contract can no longer retrieve the block hash BH(B($R_y$)+1). We mitigate this by enabling the contract to retrieve the block hash during any transaction (including customer retrieval of *V* values). Thus, even if the producer waits, other activity will enable the contract to retrieve the needed value within the period of availability. If this does not happen, the contract emits an error log and resets the block hash used to be one not yet generated. To strongly mitigate this problem for little used beacons, the contract owner should arrange for some party to access the contract at least every 256 blocks to ensure that the block hash is retrieved within the time constraints.
5. Miners (not collaborating with the producer) may try to affect $R_x$ by throwing out discovered blocks that have block hashes that will produce undesirable random numbers. However, miners must compute the block hash to be used, BH(B($R_y$)+1), prior to $V_x$ being revealed and thus this won't work. This is why the block number in which $V_x$ is processed by the contract must be at least 2 more than the block number where the last valid *V* value was processed by the contract. Note that a separate vulnerability arises if one uses the block hash of the block where the last *V* value was processed and so that was not available as an option.
6. The contract owner has only the ability to register and de-register the producer. De-registration only occurs after a set number of blocks (eliminating the possibility of the contract owner seeing a revealed $V_x$ value message and trying to remove the producer before the contract processes it). With respect to registering a producer, its first message is used only to set the initial $V_x$ Merlin value and so registration can't be used to influence or control the *V* values.
7. An attacker could compromise the producer but they would still have to produce the values on the pre-determined Merlin chain. To influence the results they would have to collaborate with a group of miners (this attack is discussed in the next section).
8. The producer who has sent some $V_x$ can predict an $R_{x+1}$ after the next block hash has been calculated. Our mitigation of this is for the contract to publish $T_{x+1}$ which indicates at what time the producer could have predicted $R_{x+1}$ (this is usually less than a minute in the past).
9. Since the producer can predict the next *R* value, it may not send some $V_x$ because revealing it will generate an $R_x$ that is deemed undesirable (e.g., the producer made a bet on the outcome). However, then it must stop producing any values because the contract will wait for $V_x$. We mitigate this by requiring producers to keep an offsite backup copy of their Merlin chain. This does not stop a producer from refusing to reveal $V_x$. However, it does eliminate their ability to claim an inability to reveal due to a hardware failure or natural disaster. This weakness could be more strongly mitigated in future work by requiring the producer to submit a timelock puzzle [17] along with each *V* value. Such puzzles would allow contract customers to perform an expensive computation on a $V_{x-1}$ to reveal any $V_x$ withheld by the producer. The producer couldn't lie at the right moment because they can't predict an $R_x$ when sending in a $V_{x-1}$ (and lying in general is easy to detect by solving the timelock puzzle).

## 5.3 Residual Security Flaw

The remaining security flaw is that the producer (or an attacker that has compromised the producer) may collaborate with a set of miners to attempt to influence, but not control, $R_x$. The malicious producer would provide the collaborating miners the value $V_x$, enabling them to compute a candidate $R_x$ if they successfully mine block $\text{B}(R_y)+1$. If this is a desirable outcome, they publish the completed block to the mining community. If not, they discard the completed block and lose the associated block reward and transaction fee funds. We mitigate this attack with our multiple producer contract.

## 6 Multiple Producer Contract

The multiple producer contract permits multiple producers to submit values to mitigate the possibility of a single producer collaborating with a group of miners. Each producer is handled independently using the single producer protocol from Sect. 5.1 (with some exceptions) and the contract maintains a beacon independently for each producer. When all beacons have pulsed, the contract pulses $R$ and $T$ values derived from the combination of beacon pulses. We call this combined output a lighthouse pulse. We change our notation to handle multiple producers as follows. We identify each producer with an integer, add this as a subscript to each variable, and let each variable refer to its most recent value. Thus, $R_1$ references the most recent $R$ value for producer 1. We use $R_L$ and $T_L$ to refer to the most recent lighthouse output.

The contract handles each producer using the single producer protocol from Sect. 5.1 with the following exceptions (that force the beacons to progress in a lockstep manner):

1. Once pulsed, beacons are not allowed to pulse again until the lighthouse pulses. If a producer sends additional messages prior to the lighthouse pulse, they are marked as invalid.
2. The '$R_y$' references in Sect. 5.1 now correspond to the $R_L$ values produced by the lighthouse (not the particular producer's beacon). This causes all beacons to use the same block hash for each beacon pulse.

Once all beacons have pulsed, the lighthouse pulses as follows:

$$R_L = R_1 \oplus R_2 \oplus \ldots \oplus R_m \tag{3}$$

where ⊕ is exclusive or (XOR) and *m* represents the number of participating beacons. This has the convenient feature that the lighthouse output using only a single producer is identical to that producer's beacon output.

$$T_L = \max(T_1, T_2, ..., T_m) \tag{4}$$

While not necessary, the lighthouse will work more efficiently if all producers synchronize their time (e.g., using the Network Time Protocol [14]) and issue messages at some agreed upon interval.

Each producer's beacon follows the single producer protocol and thus has the same security advantages. The small exceptions to the protocol in Sect. 6 do not affect the per beacon security analysis. Each beacon is still secure unless both the producer and a group of miners collude. The small exceptions cause the beacons to produce in lockstep. Due to the common block hash used, no beacon can predict the lighthouse output until after the block hash has been calculated (at which point the potentially malicious beacon has already committed to its next value).

This leaves open the possibility that a set of *t* malicious producers could collaborate on which will refuse to reveal in order to try to manipulate $2^t$ bits. However, any such activity will be publicly viewable, will cause the lighthouse to stop production, and cause the contract owner to deregister any such producers. The producers can't claim technical failures because they are required to keep a backup copy of their Merlin chains.

The only way to influence the $R_L$ values then is for all producers to collaborate with each other and also with a group of miners. They can then throw out successfully mined but undesirable blocks (those that would produce an unwanted $R_L$ value). In no situation can the $R_L$ value be controlled (i.e., directly chosen).

However, there is one remaining weakness that must be addressed. If all producers colluded when initially creating their Merlin chains then they could use the same *V* value making the beacons all pulse the same value. If there are an even number of producers, this will force $R_L$ to be 0 since it used XOR. To mitigate this, our contract simply refuses to pulse an $R_L$ value equal to 0. This obviously reduces the output state space by 1.

## 7   Empirical Work

We implemented our multiple producer contract using the Solidity language [5] and deployed it to the Ethereum test network. The test network is identical to the production network except the Ether has no real world value. Given that our system does not rely on the transfer of digital assets, the test network works just as well for our lighthouse as the real Ethereum network. We also created distributed application (DApp) software to enable producers to submit pulses to the contract and for customers to retrieve *R* values. We used multiple producers and tested the contract's ability to generate the independent beacon values as well as the lighthouse values.

We found that coding our contracts in Solidity was rather straightforward. The main challenges were that we easily ran out of gas (performed too much computation) or ran out the very limited stack space for individual functions. However, creating the beacon software that submitted pulses to the contract was much more difficult since very little documentation exists on how to enable a program outside of Ethereum to communicate with an Ethereum contract.

We didn't use the main Ethereum network for our empirical testing because the current contract execution prices made it too expensive (due to Ether currency speculation). The price of Ethereum has risen from $8.00 per Ether to $358 per Ether in six months [1] (as of June 20, 2017) and the gas fees have not dropped accordingly although Ethereum has a mechanism to do so. Table 2 shows the costs of the main functions in terms of Ether, USD on January 2017, and USD on June 2017.

If a producer pulses once a minute, the cost using June 2017 prices would be $673,000 USD per year. Using January 2017 prices, it would be $17,870 USD (which the authors believe to still be excessively high).

**Table 2.** Approximate Ether and USD Costs of Lighthouse Functions as of 2017-06-15

| Request Type | Gas | Ether | USD (2017-06-20) | USD (2017-01-01) |
|---|---|---|---|---|
| Contract Deployment | 1.9M | .0399 | $14.29 | $0.32 |
| Register Producer | 205k | .0043 | $1.54 | $0.035 |
| Producer Pulse | 200k | .0042 | $1.50 | $0.034 |
| Retrieve Output | 22k | .000462 | $0.17 | $0.0037 |

Due to these cost issues, future implementations of our contract may use an alternate to Ethereum or a private Ethereum network. This latter approach is fully supported by the Ethereum development tools and would be privately managed but publicly accessible. Another option is to design the system so that the users of the system pay the cost by charging a small fee for each delivered random number.

## 8  Conclusion

It is possible to use cryptocurrency smart contracts to create a distributed consensus protocol to publicly produce a stream of trustworthy random numbers. Our contract design eliminates both prediction and control attacks. Neither is it possible for any entity to change the values once published. What is possible is that the output might be indirectly influenced without being directly controlled but this can be mitigated by registering multiple producers.